\documentclass[12pt]{article}
\usepackage{amsmath}
\usepackage{amsfonts,amssymb,amsthm}

\newif\ifpdf
     \ifx\pdfoutput\undefined
     \pdffalse 
\else
     \pdfoutput=1 
     \pdftrue
\fi

\ifpdf
     \usepackage[pdftex]{graphicx}
     \DeclareGraphicsExtensions{.pdf, .jpg, .tif}
\else
     \usepackage{graphicx}
     \DeclareGraphicsExtensions{.eps, .jpg, .gif}
\fi

\newtheorem{thm}{Theorem}
\newtheorem{pro}[thm]{Proposition}
\newtheorem{lem}[thm]{Lemma}
\newtheorem{cor}[thm]{Corollary}
\newtheorem*{defn}{Definition}

\newcommand{\InsertFig}[4]
{\begin{figure}[ht]
       \centerline{
         \includegraphics[width=#4]{#1}
       }
       \caption{{\footnotesize  #2}}
       \label{#3}
\end{figure}}

\newcommand{\tr}{\mbox{tr}}
\renewcommand{\div}{\mbox{div}}
\newcommand{\grad}{\nabla}

\newcommand{\veps}{{\varepsilon}}
\newcommand{\R} {\mathbb{R}}
\newcommand{\Z} {\mathbb{Z}}
\newcommand{\N} {\mathbb{N}}

\newcommand{\C} {{\cal C}}

\newcommand{\cN}{{\cal N}}
\newcommand{\cP}{{\cal P}}

\newcommand{\V} {{\cal V}}
\newcommand{\W} {{\cal W}}

\newcommand{\Eq}[1]     {(\ref{#1})}
\newcommand{\Th}[1]     {Thm.~\ref{#1}}
\newcommand{\Lem}[1]    {Lem.~\ref{#1}}
\newcommand{\Cor}[1]    {Cor.~\ref{#1}}
\newcommand{\Pro}[1]    {Prop.~\ref{#1}}

\newcommand{\Def}[1]    {Defn.~\ref{#1}}
\newcommand{\Fig}[1]    {Fig.~\ref{#1}}
\newcommand{\Sec}[1]    {\S\ref{#1}}

\begin{document}

\title{Heteroclinic intersections between Invariant Circles of Volume-Preserving Maps}
\author{H. E. Lomel\'{\i}\thanks{HL wants to thank the Math
      department at UCSC for their hospitality during the academic
      year 2000--2001.  Useful conversations with R. Montgomery are
      gratefully acknowledged.  Also he would like to thank the
      support of Asociaci\'{o}n Mexicana de Cultura A.C.}\\
      Department of Mathematics\\
      Instituto Tecnol\'{o}gico Aut\'{o}nomo de M\'{e}xico\\
      M\'{e}xico DF 01000
\and J. D. Meiss\thanks{
        Support from NSF grant DMS-0202032 gratefully acknowledged}\\
        Department of Applied Mathematics\\
        University of Colorado\\
        Boulder CO 80309-0526}
\maketitle
\begin{abstract}
We develop a Melnikov method for volume-preserving maps with codimension one
invariant manifolds. The Melnikov function is shown to be related to the flux 
of the perturbation through the unperturbed invariant surface. As an example,
we compute the Melnikov function for a perturbation of a three-dimensional map 
that has a heteroclinic connection between a pair of invariant circles. The
intersection curves of the manifolds are shown to undergo bifurcations in homology.
\subsection*{AMS classification scheme numbers:}
34C20,34C35,34C37,58F05,70H99
\end{abstract}

%
%
\section{Introduction}

Volume-preserving maps on $\R^{3}$ provide an interesting and
nontrivial class of dynamical systems and give perhaps the simplest,
natural generalization to higher dimensions of the much-studied class
of area-preserving maps.  They also arise in a number of applications
such as the study of the motion of Lagrangian tracers in
incompressible fluids or of the structure of magnetic field lines
\cite{Holmes84, Lau92}.  Experimental methods have only recently been
developed that allow the visualization of particle trajectories in
three-dimensional fluids \cite{Mezic02, Shinbrot01}. 
The infinite dimensional group of volume-preserving diffeomorphisms is also at the 
core of the ambitious program to reformulate hydrodynamics \cite{Arnold}.

While some of the results for area-preserving maps generalize to the
volume-preserving case, the study of transport in such systems is
still in its infancy \cite{Piro88, Feingold89,MacKay94}.  The theory
of transport is based on dividing phase space into regions separated
by partial barriers through which flux can be measured.  For the
area-preserving case, the natural partial barriers are formed from the
stable and unstable manifolds of periodic orbits or cantori
\cite{MMP84, Meiss92}.  Primary intersections can be used to form
\emph{resonance zones} \cite{MMP87, Easton91}---regions of phase space
that are bounded by alternating stable and unstable segments joined at
primary intersection points.  Because the intersection points are
primary, a resonance zone is bounded by a Jordan curve and has an
exit and an entry set \cite{Easton93}.  The area of each of these sets
is the geometric flux, the area leaving the resonance zone each
iteration of the map.  The images of the exit and entry sets and their
intersections completely define the transport properties of the
resonance zone \cite{RomKedar88}.

Thus the beginning of a generalization of this theory to higher
dimensions is the study the intersections of codimension-one stable
and unstable manifolds for volume-preserving maps.

As is well-known, a transversal intersection of stable and unstable
manifolds is associated with the onset of chaos, giving rise to the
construction of Smale horseshoes.  A widely used technique for
detecting such intersections is the Melnikov method.  Given
a system with a pair of saddles, and a heteroclinic or saddle
connection between them, the Melnikov function computes rate at which
of change the distance between the manifolds changes with a perturbation.  
The integral of the Melnikov function between two neighboring primary intersection
points is the first order term in the geometric flux \cite{MM88,
Kaper91}.

Most applications of the method are for two-dimensional maps and flows
\cite{Easton84, Delshams96, Delshams97, Lomeli97}, though a Melnikov
method for a three-dimensional incompressible flow was developed in
\cite{Mezic94}.  In this latter case the perturbation may be
periodically time dependent, and the Poincar\'{e} map of the system is
assumed to have a hyperbolic invariant curve, with two-dimensional
manifolds.

For the case of maps, the analogue of Melnikov integral is an infinite
sum whose domain is the unperturbed connection.  As usual, a simple
zero of this function corresponds to a transverse intersection of the
manifolds for the perturbed map.  We developed a Melnikov method for
three-dimensional maps in \cite{Lomeli00} to study and classify
intersections of stable and unstable manifolds for fixed points.

In this paper we generalize this method to the problem of
detecting heteroclinic orbits between a pair of normally hyperbolic
invariant sets in volume-preserving maps on $\R^{n}$.  Our application
is to the case of invariant circles for a three-dimensional map.

To obtain a Melnikov function, we must define an appropriate measure
of the distance between a manifold and its perturbation.  In
\cite{Lomeli00} we used the cross product of pair of tangent vectors
fields to obtain this distance.  Different versions of Melnikov's
method have used other ways of measuring the splitting between the
unperturbed separatrix and the perturbed one, though naturally only
the normal distance is well-defined in the codimension-one case.  This
is appropriately measured using an{\em adapted normal} vector field or
differential form.  An adapted normal is a normal field to the saddle
connection that is invariant under the dynamics.  If the map is
integrable, then the gradient of an integral can be used to construct
the adapted normal, but the concept applies more generally to
nonintegrable systems.

We use the Melnikov function to construct a {\em flux-form}, an $(n-1)$-form
whose integral over a {\em fundamental domain} on the connection measures
the first order flux through the connection.  The fundamental domain
is an annulus that generates the entire manifold upon iteration. 
Since the map is volume-preserving, the net (algebraic) flux always
vanishes, but the one-way (geometric) flux gives a measure of the
transport.

In \cite{Lomeli00} we introduced a family of volume-preserving maps
that have a saddle connection between a pair of fixed points.  This
family is obtained from a family of planar twist maps with a saddle
connection \cite{Lomeli96}.  This family can be modified so that it
has a pair of invariant circles with a corresponding
of saddle connection.  We perturb this family by composing it with a
near-identity, volume-preserving map, thus producing examples of
volume-preserving maps with transverse heteroclinic orbits.

We study the curves of zeros of the Melnikov function on a fundamental
domain of the unperturbed manifold.  Using the map to identify the two
boundaries, the fundamental domain becomes a torus.  Thus the zeros of the
Melnikov function can be characterized by their homology on this
torus.  We show that as the parameters of the map are varied, the
homology of these curves undergoes bifurcations, and that these
bifurcations strongly influence the geometric flux.

%
%
\section{Basic definitions and properties}

Suppose $f_0:\R^{n}\rightarrow\R^{n}$ is a diffeomorphism on 
$n$-dimensional Euclidean space.  A smooth perturbation of $f_{0}$ is
a family of functions $f_{\veps} \equiv f(\cdot ,\veps)$ such that
$f(\cdot,0)=f_{0}$ and $f(x,\veps)$ is smooth in both variables.
We now define a vector field on $\R^{n}$ that will be
used to measure the motion of an invariant manifold.

\begin{defn}[Perturbation vector field]\label{def:X_epsilon} 
	Given a perturbation $f_{\veps}$ of $f_{0}$, define
	the vector field $X_{\veps}$ for any point $x\in \R^{n}$ by
	\begin{equation}\label{eq:pertvect}
	      X_{\veps}(x)\equiv  \left[\frac{\partial }{\partial \veps}
	        f_{\veps}(y)\right]_{y=f_{\veps}^{-1}(x)}\;.
	\end{equation}
\end{defn}
Perturbation vector fields have some special properties. First, it is easy
to see that $X_{\veps}$ is independent of $f_{0}.$ Second, if one regards
$X_{\veps}$ as a time dependent vector field (where time is $\veps$), then
$y(\veps) = f_{\veps}(x)$ is the solution of the initial value problem
\[
     \frac{dy }{d \veps}\;\equiv X_{\veps}(y) \;, \quad y(0)=f_{0}(x) \;.
\]
Thus if we let $F_{t,s}=f_{t}\circ f_{s}^{-1},$ then $F$ represents
the flow of the nonautonomous vector field $X_{\veps}$ \cite[Thm. 
2.2.23]{Abraham}.  Since $F$ is volume-preserving, the vector
field $X_{\veps}$ has zero divergence with respect to $\Omega$
\cite[Thm.  2.2.24]{Abraham}.

It is often convenient to define a perturbed family by composing
$f_{0}$ with an $\veps$ dependent perturbation:
\begin{equation}\label{eq:pert}
       f_{\veps} = (id+\veps P_{\veps}) \circ f_{0} =f_0+\veps
P_{\veps}\circ f_{0} \;.
\end{equation}
In this case, the vector field is the first order approximation to
the perturbation, $X_{0}(x) = P_{0}(x)$.

%
%

\subsection{Invariant manifolds}

Suppose the family $f_\veps$ has a family of invariant manifolds
$\W_\veps \hookrightarrow \R^{n}$.  In this paper, we will assume that
$\W_\veps$ is a codimension-one surface.  Our goal is to understand,
at least to first order, the relation between the perturbation vector
field and the way these invariant manifolds evolve with $\veps$. 
Later, we will restrict ourselves to the case in which $\W_{\veps}$
consists of pieces of stable and unstable manifolds of some invariant
set.  When $\W_{\veps}$ is a smooth graph over $\W_0$, we can define a
map that is adapted to the $\veps$ parameterization:

\begin{defn}[Adapted deformation] \label{def:adaptDef}
       A map $\phi:\W_0 \times(-\veps_{0},\veps_{0})\rightarrow\R^{n}$, is
       \emph{adapted} to $\W_{\veps}$, if there is an $\veps_{0}>0$ such
       that
       \begin{itemize}
     \item $\phi_{\veps}=\phi(\cdot,\veps)$ is a diffeomorphism
         $\phi _{\veps}:\W_0\rightarrow\W_{\veps} \;,\;
         \forall\veps\in(-\veps_{0},\veps_{0})$.
     \item $ \phi_0=\phi(\cdot,0)=id_{\W_0}$.
       \end{itemize}
\end{defn}

There is quite a bit of freedom in the choice of $\phi$; however, only
the normal behavior is important for our application, since it
measures the actual motion of $\W_\veps$ with $\veps$, and this is
unique:
\begin{pro} \label{pro:twoadapted}
       Suppose $\phi_{\veps}$ and $\tilde{\phi}_{\veps}$ are two adapted
       deformations for a family of invariant manifolds $\W_{\veps}$. Then
       \[
     \left[
        \frac{\partial}{\partial\veps} \tilde \phi_{\veps}(x) -
        \frac{\partial}{\partial\veps}            \phi_{\veps}(x)
           \right]_{\veps=0}
     \in T_{x}\W_0.
       \]
\end{pro}

\proof Since both $\phi_{\veps}(x)$ and $\tilde\phi_{\veps}(x)$ are
points in $\W_{\veps}$, the curve $C_{\veps}(x) =
\phi^{-1}_{\veps}(\tilde\phi_{\veps}(x))$ is a curve in $\W_0$
parameterized by $\veps$, and $C_{0}(x) = x$.  Thus its derivative at
zero is a tangent vector to $\W_0$:
\[
      \left[ \frac{\partial}{\partial \veps} C_{\veps}\right]_{\veps=0}
            \in T_{x}\W_0.
\]
Thus
\[
       \left. \frac{\partial}{\partial\veps}
\phi_\veps({C_\veps(x))}\right|_{\veps=0}  =
        \left. \frac{\partial}{\partial\veps} \phi_{\veps}(x)  \right|_{\veps=0}
           +D\phi_0(x)
\left[\frac{\partial}{\partial\veps}C_{\veps}(x)\right]_{\veps=0}=
       \left. \frac{\partial}{\partial\veps} \tilde\phi_{\veps}(x)
\right|_{\veps=0}  \;,
\]
Since $D\phi_0(x) = I$, this gives the promised result.
\qed

We will use this proposition to compute the Melnikov function in
\Sec{sec:melnikov}.  For this we need to measure rate of change of an
invariant manifold with respect to a perturbation---changes in the
tangent direction are unimportant.  In order to measure the change in
the normal direction, we introduce the concepts of adapted normal
vector fields and adapted forms.

%
%
\subsection{Adapted normals}

We will measure the splitting by using a normal to the invariant
manifold $\W_0$, that throughout this paper will be a
codimension-one submanifold, i.e. a surface.  To be useful, the
normal field should evolve in a precise way under the unperturbed
map or, as we say, be ``adapted'' to the dynamics.

First we recall some notation.  Let $v$ be a vector field and $f$
a diffeomorphism.  The pull-back of $v$ under $f$ is $(f^{*}v)(x)
= (Df(x))^{-1}v(f(x))$.  Similarly, the pull-back of a $k$-form
$\omega$ is $(f^* \omega)_x(v_1,v_2,...v_k) =
\omega_{f(x)}(Df(x)v_1,\ldots, Df(x)v_k)$.  Finally the inner
product of a vector field $Y$ with a $k$-form $\omega$ is defined
as the $(k-1)$-form $i_Y \omega = \omega(Y,\cdot,\ldots,\cdot)$.

\begin{defn}[Adapted normal field] \label{def:adaptedVect}
      Suppose that $f:\R^{n}\rightarrow\R^{n}$ is a diffeomorphism with
      an invariant surface $\W$, and there is given an inner product
      $\langle , \rangle $ for vectors on $\R^{n}$.  An adapted normal
      field is a smooth function $\eta:\W\rightarrow\R^{n}$ such that
      \begin{itemize}
     \item $\eta(x )\neq 0$ for all $x \in\W$.
     \item $\eta(x )$ is normal to the surface for all $x \in\W$, that
           is $\eta(x )\in T_{x }\W^{\perp}$.
     \item  For all vector fields $Y:\W\rightarrow\R^{n}$ we have that
         \begin{equation} \label{eq:adaptVect}
         f^{*}\langle \eta,Y\rangle =\langle \eta,f^{*}Y\rangle .
         \end{equation}
      \end{itemize}
\end{defn}
The geometry is shown in \Fig{fig:adaptedNormal}.  Note that the
pullback of a scalar function $g:\R^n\rightarrow \R$ is $f^*g(x) =
g(f(x))$; thus if we define $Z(x) = f^*Y(x)$, \Eq{eq:adaptVect} is
equivalent to
\[
\langle \eta(x), Z(x) \rangle = \langle \eta(f(x)), Df(x)Z(x) \rangle
\]

\InsertFig{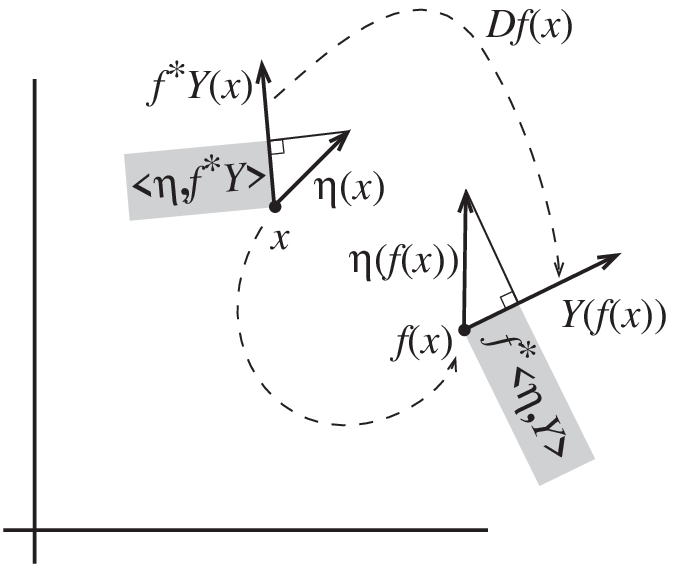}
       {$\eta$ is an adapted normal if the two rectangles shown have the same
       area.}
       {fig:adaptedNormal}{3in}

Adapted normals can be thought of as a generalization of the gradient
normal that one gets from a first integral.  Recall that the gradient
of a smooth function $J:\R^{n} \rightarrow\R$ is the unique vector
field, $\grad J$, such that for all vector fields $Y$ on $\R^{n}$,
\begin{equation}\label{eq:contract}
      i_{Y}dJ \equiv dJ(Y) =\langle \grad J,Y\rangle .
\end{equation}
If $f$ has a nondegenerate first integral, $J=J\circ f$, then
\Eq{eq:contract} implies that $f^{*}\langle \grad J,Y\rangle
=\langle \grad (J\circ f),f^{*}Y\rangle $.  Therefore, if the
diffeomorphism $f$ has a first integral $J$, then $\grad J$ is an
adapted vector field, provided it doesn't vanish on $\W$.

If we are using the standard inner product on $\R^{n}$, then we can
characterize adapted normals more concretely.
\begin{pro}
      Let $\eta:\W\rightarrow\R^{n}$ be a smooth function defined on the
      invariant surface $\W$, and suppose $\langle u,v \rangle = u^t
      \cdot v$ is the standard inner product on $\R^n$.  Then $\eta$
      satisfies \Eq{eq:adaptVect} for all vector fields
      $Y:\W\rightarrow\R^{n}$ if and only if, for all $x \in\W$
      \[
      Df(x )^{t}\eta(f(x ))=\eta(x ).
      \]
\end{pro}

In the general case $\W$ is not defined as the level set of an
invariant, and it is not easy to show that an adapted normal field
exists.  However, when the map is volume-preserving and we are given
an appropriate parameterization of the invariant surface, an adapted
normal vector field can easily be constructed.

\begin{lem} \label{lem:VectConstruct}
      Suppose that $f:\R^{n}\rightarrow\R^{n}$ preserves the volume-form
      $\Omega$, and has a smooth invariant surface $\W$.  Suppose
      $k:\R^{n-1}\rightarrow\R^{n}$ is a nondegenerate
      parameterization of $\W$ with the property
      \begin{equation} \label{eq:param}
     f(k(u))=k(u+\delta) \;,
      \end{equation}
      for a constant $\delta \in \R^{n-1}$.  Then the vector field
      $\eta:\W\rightarrow \R^{n}$ restricted
      to $\W$  defined by
      \begin{equation}\label{eq:paramvector}
         \langle \eta, \cdot \rangle = \Omega(\cdot,
\partial_{u_1}k,\partial_{u_2}k,\ldots,\partial_{u_{n-1}}k)
      \end{equation}
      is an adapted vector field on $\W$.
\end{lem}

\proof The nondegeneracy of the parameterization implies that $\eta
\neq 0$ on the surface $\W$. Condition \Eq{eq:param} implies that
\[
      D_{u}k(u+\delta)  = D_{u}\left(f\circ k\right)(u) =
             D_{x}f\left( k(u)\right) D_{u}k(u) \; ;
\]
therefore, since $f^{*}\Omega = \Omega$, we have
\begin{align*}
    f^* \langle \eta(k) ,Y(k)\rangle
      & = \langle \eta(f(k)) ,Y(f(k))\rangle \\
      & = \Omega_{f(k)}\left(Y(f(k)),
\partial_{u_1}k(u+\delta),\ldots,\partial_{u_{n-1}}k(u+\delta)\right)
\\
      & = \Omega_{f(k)}\left(Y(f(k)),
          Df(k)\partial_{u_1}k(u),\ldots,Df(k)\partial_{u_{n-1}}k(u)\right)
\\
      & = \Omega_{x}\left(Df(x)^{-1}Y(f(k)),
          \partial_{u_1}k(u),\ldots,\partial_{u_{n-1}}k(u)\right) \\
      & = \langle \eta(k),f^{*}Y)\rangle
\;. \qed
\end{align*}

%
%
\subsection{Adapted one-forms}

An alternative concept to that of adapted vectors are adapted
forms.  The advantage of one versus the other approach is mainly a
question of taste, though differential forms can be used without
assuming an inner product. We now define an ``adapted one-form.''

\begin{defn}[Adapted one-form]\label{defn:adapted}
       Suppose that $f:\R^{n}\rightarrow\R^{n}$ is a
       diffeomorphism. An adapted one-form on an invariant surface
       $\W$ is a smooth function $\nu: T_{\W}\R^{n} \rightarrow \R$
       such that
       \begin{itemize}
     \item $\nu_{x}$ is nondegenerate for all $x\in\W$.
     \item $\nu_{x}(v) = 0$ for all $v\in T_{x}\W$.
     \item $f^{*} \nu = \nu$
       \end{itemize}
\end{defn}

Note that when $\nu$ is an adapted one-form then for each $x\in
\W$, $\ker(\nu_x)=T_x\W$, but that since $\nu$ is nondegenerate,
it will not be zero for vectors that are not tangent to $\W$. As
before, we note that if $\W$ is given as the level surface of an
invariant function $J$, i.e. if $J(f(x)) = J(x)$, then an adapted
one-form is easy to obtain: the one-form $dJ$ is adapted provided
only that $J$ has no critical points on $\W$.  This follows
because $f^{*}dJ = d(J \circ f) = dJ$.

Given an inner product, $\langle \cdot, \cdot \rangle $ we can
always associate a unique vector field, $\eta$ with a form $\nu$,
through $i_X\nu=\nu(X) =\langle \eta,X\rangle .$ Here $\eta:\W\to\R^{n}$ is a
smooth function.  It is easy to see that, if the $\eta$ is adapted,
then $\nu$ is also adapted.  Conversely, given an adapted one-form, we
can find an adapted normal field, through the same relation.

\begin{pro} \label{pro:vectform}
      Let $\eta$ and $\nu$ be related through
      \begin{equation}\label{eq:form_vector}
     i_X\nu=\langle \eta,X\rangle  \;.
      \end{equation}
      Then $\eta$ is an adapted normal if and only if $\nu$ is an
      adapted form.
\end{pro}

Using this with \Lem{lem:VectConstruct} implies immediately.

\begin{cor} \label{lem:FormConstruct}
       Let $f:\R^{n}\rightarrow\R^{n}$ preserve a volume-form $\Omega$,
       and $\W$ be a smooth invariant surface.  Suppose $k
       :\R^{n-1}\rightarrow\R^{n}$ is a nondegenerate parameterization of
       $\W$ such that $f(k(u,v))=k(u+\delta)$ for constant $\delta \in
       \R^{n-1}$.  Then
      \begin{equation}\label{eq:paramform}
     \nu = \Omega(\cdot, \partial_{u_{1}}k, \ldots,
\partial_{u_{n-1}}k)
      \end{equation}
      is an adapted one-form on
      $\W$.
\end{cor}

%
%
\subsection{Example}

Let $f:\R^{3}\rightarrow\R^{3}$ be the two parameter family of
diffeomorphisms
\[
       f(x,y,z)=\left(\begin{array}[c]{c}
     e^{\tau} (  x \cos\theta - y \sin\theta) \\
     e^{\tau} (  x \sin\theta + y \cos\theta) \\
     \frac12(x^{2} + y^{2}) + e^{-2\tau} z
       \end{array}\right)  \;,
\]
where $\tau$ and $\theta$ are constants. It is easy to see that $f$
preserves the standard volume-form $\Omega=dx\wedge dy\wedge dz$.
In addition, $f$ has an invariant surface given by
\[
       \W=\left\{ (x,y,z)\in\R^{n}\setminus \{0\}:
            x^{2}+y^{2} =  4z \sinh(2\tau) \right\} \; ;
\]
however, the ``obvious'' function $x^{2}+y^{2}-4 z\sinh(2\tau) $ is
{\em not} invariant.
Instead, we parameterize $\W $ with the function
$k:\R^{2}\rightarrow\R^{3}$ given by
\[
       k(u,v)=\left(\begin{array}[c]{c}
             e^{u\,\tau} \cos v\\
             e^{u\,\tau} \sin v\\
             (4\sinh2\tau)^{-1} e^{2\,u\,\tau}
                   \end{array}\right)  \;.
\]
The function $k$ is nondegenerate and
satisfies\[
       f(k(u,v))=k(u+1,v+\theta).
\]
so that \Lem{lem:VectConstruct} and \Lem{lem:FormConstruct} apply.
Using \Eq{eq:paramform}, we find
that $ \nu= -2\tau z ( xdx + ydy - 2 \sinh(2\tau) dz) $ an adapted
form on $\W$.  In other words $f^{*}\nu = \nu$ and $\ker
\nu_p=T_p\W$, as can be explicitly verified.  Also using
\Eq{eq:paramvector}, it is possible to show that
\[
      \eta(x,y,z)=\left(
     \begin{array}[c]{c}
         -2\tau x z \\
         -2\tau y z\\
         4\tau z \sinh 2 \tau
     \end{array} \right)
\]
is an adapted normal field on $\W $.  In other words, it satisfies
$Df(x)^{t}\eta (f(x))=\eta(x)$, for each $x\in\W$.  Note that $\nu$
and $\eta$ are related through $i_X\nu=\langle \eta,X\rangle $.

%
%
\section{Melnikov Function} \label{sec:melnikov}

Suppose that the diffeomorphism $f_0$ has two normally hyperbolic
invariant sets $p$ and $q$, and a codimension-one surface $\W = W^u(p)
= W^s(q)$ that is a saddle connection between them.  Upon
perturbation, suppose that the corresponding invariant sets $p_\veps$
and $q_\veps$ of $f_\veps$ have a stable manifold $\W_{\veps}^{s}$ and
unstable manifold $\W_{\veps}^{u}$.  Then in our notation, the
classical Melnikov function is the smooth function $M_{\nu}:\W
\rightarrow\R$ on the saddle connection $\W$ defined by
\begin{equation}\label{eq:melnikov}
        M_{\nu} \equiv  \nu \left(
                  \left. \frac{\partial}{\partial\veps}\right| _{\veps=0}
         \left(\phi_{\veps}^{u}-\phi_{\veps}^{s}\right)
         \right)  \;,
\end{equation}
for a given adapted form $\nu$ on $\W$, and a given pair of adapted
perturbations $\phi_{\veps}^{s}$ and $\phi_{\veps}^{u}$ corresponding
to the stable and unstable manifolds, respectively.  Thus $M_\nu$
measures the relative ``velocity" of the manifolds as a function of
$\veps$.  While $M_\nu$ appears to depend on the choice of adapted
perturbations, we will show that it does not.

A function similar to $M_{\nu}$ was used in \cite{Lomeli00}, to study the
topology of heteroclinic connections of fixed points. Our purpose is to apply
the method to the case of invariant circles, as illustrated in
\Fig{fig:saddle}.

\InsertFig{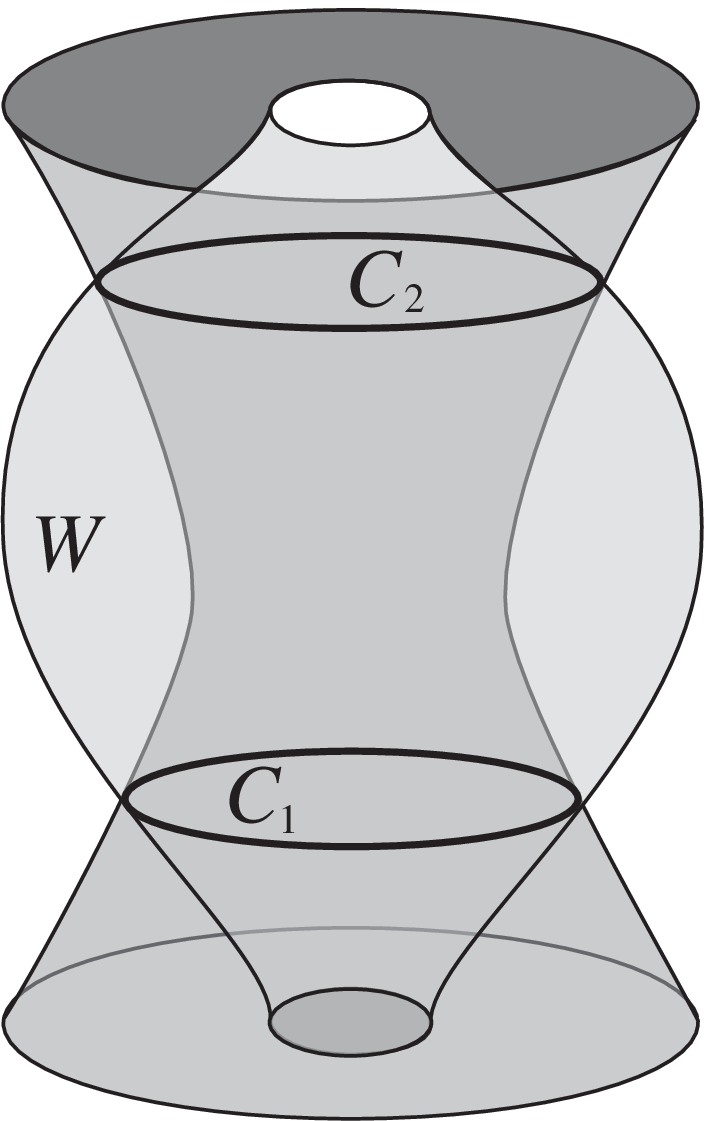}
       {Two normally hyperbolic invariant circles $C_{1}$ and $C_{2}$
       with a saddle connection $\W$.}
       {fig:saddle}{2in}

%
%
\subsection{The fundamental iterative relation}

The fundamental relation used in deriving the Melnikov function is an
iteration formula obtained by combining the definition of adapted 
one-form and \Pro{pro:twoadapted}.

\begin{thm}\label{thm:iterative}
       Suppose $f_{\veps}$ is a family of diffeomorphisms with invariant
       surfaces $\W_{\veps}$.  Let $\nu$ be an adapted one-form and
       $\phi_{\veps }$ be an adapted deformation on $\W_{0}$.  Define
       $\mu:\W\to \R$ by
       \[
              \mu(x) = \nu \left(\left.  \partial_{\veps}
              \phi(x,\veps)\right|_{\veps=0}\right) \;.
       \]
       Then
       \begin{equation}\label{eq:iterative}
                 \mu-\mu \circ f_{0}^{-1}= \nu(X_{0}) \;,
       \end{equation}
	   where $X_0$ is the perturbation vectorfield, \Eq{eq:pertvect}.
       Moreover, if $\tilde{\phi}_{\veps}$ is another adapted
       deformation and $\tilde \mu$ is defined similarly to $\mu$, then
       $\mu = \tilde \mu$.
\end{thm}

\proof By \Pro{pro:twoadapted} the difference between the derivatives
of two adapted diffeomorphisms is tangent to $\W$, and by
\Def{defn:adapted}, $\nu$ vanishes on any tangent vector.  Thus $\mu =
\tilde \mu$.  To compute the second relation, use $f_{0}^{*}\nu =
\nu$ to find
\begin{align*}
       \mu  &= f_{0}^{*}\nu_{x}(\partial_{\veps}\phi_{\veps}(x)) \\
            &= \nu_{f_{0}(x)}\left(Df_{0}(x) \partial_{\veps}
\phi_{\veps}(x)\right)  \\
            &= \nu_{f_{0}(x)}\left(\partial_{\veps}f_{\veps}(\phi_{\veps}(x)) -
                         \partial_{\veps}f_{\veps}(x)\right) \;,
\end{align*}
where we suppress the $\veps = 0$ expressions for simplicity.
Using \Eq{eq:pertvect} $X_{0} = \partial_{\veps }f_{\veps }(
f_0^{-1}(x) )$, we have $\mu \circ f_{0}^{-1} = \nu_{x}\left(
\partial_{\veps}f_{\veps}(\phi_{\veps}(f_{0}^{-1}(x))) \right) -
\nu_{x}(X_{0})$, and therefore
\[
       \mu - \mu \circ f_{0}^{-1} = \nu \left( \partial_{\veps}\phi -
               \partial_{\veps}f_{\veps}(\phi_{\veps}(f_{0}^{-1}(x))) \right) +
               \nu(X_{0}) \;.
\]
Noting that $\tilde{\phi}_{\veps}= f_{\veps}(\phi_{\veps}(f_{0}^{-1}(x))$ is
also an adapted diffeomorphism, we see that the first term vanishes by
\Pro{pro:twoadapted}.  \qed

Equation \Eq{eq:iterative} gives us a recursive formula to compute the
normal component of the change in the manifold $\W_{\veps}$.

\begin{cor} \label{cor:iterative}
       Under the assumptions of \Th{thm:iterative}, for all $n\in\N$
      \[
        \mu=\mu\circ f_{0}^{-n}+\sum_{k=0}^{n-1}\nu(X_{0}) \circ f_{0}^{-k}
        \;.
       \]
       In addition, if $\displaystyle\lim_{n\rightarrow\infty}\mu\circ
       f_{0}^{-n}(x) = 0$, then
       \begin{equation}\label{eq:sum}
     \mu(x)=\sum_{k=0}^{\infty}\nu(X_{0}) \circ f_{0}^{-k}
=\sum_{k=0}^{\infty}\nu\left((f_{0}^{-k})^{*}X_{0}\right)
     \;.
       \end{equation}
\end{cor}

   These statements can be directly transcribed for adapted normals
using \Pro{pro:vectform}.

%
%
\subsection{Transversal intersections}
According to \Th{thm:iterative} and \Cor{cor:iterative}, we can
compute the Melnikov function \Eq{eq:melnikov} in terms of the first
order perturbation vector field $X_{0}$.

\begin{pro}
      Suppose $f$ has a codimension-one saddle connection $\W$ between
      two normally hyperbolic invariant sets $p$ and $q$.  Assume that
      for all $x \in p \cup q$, the perturbation vector field $X_{0}(x)
      = 0$.  Let $\nu$ be an adapted form and $\eta$ the corresponding
      adapted normal defined on $\W$.  Define the Melnikov function by
      \begin{equation}\label{eq:melnikovsum}
       M_{\nu}=\sum_{k=-\infty}^{\infty} \nu(X_{0}) \circ f^{k}_0
              = \sum_{k=-\infty}^{\infty} \langle \eta, X_{0} \rangle
\circ f^{k}_0 \;.
      \end{equation}
      Then if a point $x_{0}\in\W$ is a nondegenerate zero of
      $M_{\nu}$, the stable and unstable manifolds $W^{u}(q,f_{\veps})$ and
      $W^{s}(p,f_{\veps})$ intersect transversally near $x_{0}$ for
      $\veps$ small enough.
\end{pro}

\proof For each point $x$ in the saddle connection $\W$, there is a
neighborhood $\cN_0 \subset \W$, such that all the iterates
$f^k(\cN_0)$ are disjoint.  Moreover, since $p$ and $q$ are normally
hyperbolic, the stable manifold theorem implies that there is an
$\veps_0>0$ such that there exist adapted deformations $\phi^{u}:
{\cN_0}\times(-\veps_0,\veps_0)\to W^{u}(q,f_{\veps})$, and $\phi^{s}:
{\cN_0}\times(-\veps_0,\veps_0)\to W^{s}(p,f_{\veps})$.

Consider first the unstable part.  Let $\V = \bigcup_{k=0}^{\infty}
f_0^{-k}({\cN_0})$.  Clearly $\V$ is a immersed manifold.  Moreover, we
can extend the domain of $\phi^{u}$ to all of $\V$, by defining
\[
     \phi^{u}(x,\veps)=f_\veps^{-k}(\phi^{u}(f_0^{k}(x),\veps))\;,
\]
provided that $x\in f_0^{-k}(\cN_0)$.  It is clear that for each
$\veps\in (-\veps_0,\veps_0)$ and $x\in \V$, we have that
$\phi^{u}(x,\veps)\in W^u(q,f_\veps)$.

For each $x$, we are interested in estimating $\phi(x,\veps)$ to
first order in $\veps$.  Using $\phi^{u}$ in \Cor{cor:iterative} gives
\Eq{eq:sum} providing $\mu^{u} \circ f^{-n}(x) \rightarrow 0$.  This
is the case because $\phi^{u}(f^{-n}(x),\veps) \rightarrow 0$ so
that $\partial_{\veps} \phi^{u}$ is bounded, and
$\nu_{f_{0}^{-n}(x)} \rightarrow 0$ since it is an adapted form.

Similar analysis applies to the stable adapted deformation, and again
\Cor{cor:iterative} applies, though we iterate in the opposite
direction, to obtain $\mu^{s}= -\sum_{k=1}^{\infty}\nu(X_{0}) \circ
f_{0}^{k}$.  According to \Eq{eq:melnikov}, the difference between
$\mu^{u}$ and $\mu^{s}$ gives the Melnikov function, which yields
\Eq{eq:melnikovsum}.

Following a standard Melnikov argument based on the implicit function
theorem \cite{Wiggins88}, we conclude that if $x_0$ is a
nondegenerate zero $M_{\nu}$ then near $x_0$, the two manifolds
$W^u(f_\veps)$ and $W^s(f_\veps)$ intersect transversely.
\qed

%
%

\section{Flux}\label{sec:flux}

The flux across a surface is the volume that crosses the surface each
iterate of a map; it is an important measure of transport.
Recall that for area-preserving maps, the Melnikov
function is a measure of the distance between the stable and unstable
manifolds, and that its integral between two successive zeros is the
geometric flux that crosses the ``separatrix'' each iteration of the
map \cite{MM88,Kaper91}. The outgoing flux is exactly balanced by an 
ingoing flux, so that the net, or algebraic, flux crossing the separatrix is zero.

Here we will obtain an analogous formula for volume-preserving maps
(see also \cite{MacKay94}).  We start by constructing a flux form on 
an invariant set.  We will see that the algebraic flux
crossing the separatrix is zero.  This implies, for example that the
Melnikov function has zeros in the separatrix.

%
%
\subsection{Flux Form}
It is well known that a volume-preserving map with an invariant $J$
can be restricted to a measure preserving map on any surface $J = c$
on which $\grad J$ is nonzero.  That is, the form $\omega = {|\grad
J|^{-2}}i_{\grad J}\Omega$ is an invariant $n-1$ form for the map
$\left.f\right|_{J=c}$.  We show here that a similar preserved measure
also exists if we can find an adapted normal for an invariant surface
$\W$. We will then use this to construct a flux form on $\W$.

As usual, we assume that $f$ is a diffeomorphism with an invariant
volume-form $\Omega$, $\W$ is an invariant codimension-one
hypersurface, and $\langle\, , \,\rangle$ is an inner product on
$\R^{n}$.

\begin{pro}\label{thm:2form}
      Suppose $\eta$ is an adapted normal field on $\W$.  Then
      \[
     \omega_{\eta}=\frac{i_{\eta}\Omega}{\langle \eta,\eta\rangle }
      \]
      is a nondegenerate $(n-1)$-form on $\W$ that is invariant under
      the restricted map $\left.  f\right|_{\W}:\W \rightarrow\W$.
\end{pro}

\proof It is clear that $\omega_{\eta}$ is a nondegenerate $(n-1)$-form on
$\W$.  We need to show that $f^{*}_0\omega_{\eta}=\omega_{\eta}$ for vectors
in $T\W$.  
With some manipulations we have
\begin{align*}
      f^{*}\omega_{\eta}-\omega_{\eta}  &=\frac{i_{f^{*}\eta}f^{*}\Omega}
           {f^{*}\langle \eta,\eta\rangle }-
\frac{i_\eta\Omega}{\langle \eta,\eta\rangle }\\
      &=\frac{i_{f^{*}\eta}\Omega}{\langle \eta,f^{*}\eta\rangle
}-\frac{i_\eta\Omega
      }{\langle \eta,\eta\rangle }=i_{v}\Omega.
\end{align*}
where we define the vector field:
\[
      v=\frac{f^{*}\eta}{\langle \eta,f^{*}\eta\rangle }-
              \frac{\eta}{\langle \eta,\eta\rangle } \;.
\]
Since  $\langle \eta,v\rangle =0$ on $\W$ for each point $x\in\W$,
and $\eta$ defines the normal direction, then $v(x)\in T_{x}\W$.
Since $v$ is tangent to $\W$, the form $i_{v}\Omega$
restricted to $\W$ has to vanish.  Thus, we conclude that
$\left(f^{*}\omega_{\eta}-\omega_{\eta }\right)_{\W}=0$.  \qed

 From now on assume that $\W$ has an adapted vector field $\eta$, and the
Melnikov sum \Eq{eq:melnikovsum} exists.  We then define
\begin{defn}[Flux Form]\label{def:flux}
      The flux form  $\Phi \equiv M_{\eta}\omega_{\eta}$
	is an $(n-1)$-form on $\W$.
\end{defn}
Note that $\Phi$ might be degenerate (it has zeros), and since the space
of $(n-1)$-forms on the $(n-1)$-dimensional manifold $\W$ is
one-dimensional, it might not be unique. However, this is not the case:
\begin{lem}\label{lem:independent}
   If the Melnikov function exists, then the form  $\Phi$ is
   independent of the choice of $\eta$.
\end{lem}

\proof
The projection of $X_{0}$ onto $T\W$ is the vector
$v = X_{0}-\frac{\langle \eta,X_{0}\rangle }{\langle \eta,\eta\rangle}\eta$.
Note that
\[
     i_{X_{0}}\Omega -\langle \eta,X_{0}\rangle \omega_{\eta}= i_{v } \Omega.
\]
Since $v \in T\W$, we conclude that $i_{v } \Omega = 0$, as a
$(n-1)$-form in the surface
$\W$, and so we have
\[
       \langle \eta,X_{0}\rangle \omega_{\eta} =  i_{X_{0}}\Omega
\]
for all vectors in $T\W$. This implies that the summand of
$M_{\eta}\omega_{\eta}$, recall \Eq{eq:melnikovsum}, can be rewritten
\[
    (f_{0}^{k})^{*}\langle \eta, X_{0}\rangle\omega_{\eta}
           = (f_{0}^{k})^{*} i_{X_{0}}\Omega \;,
\]
which is independent of $\eta$.\qed

Thus we have
\begin{equation}\label{eq:fluxform}
     \Phi = M_{\eta}\omega_{\eta} = \sum_{k=-\infty}^{\infty} (f_{0}^{k})^{*} i_{X_{0}}\Omega \;.
\end{equation}
Since $\omega_{\eta}$ is  nondegenerate, the degenerate points of the flux form correspond to zeros of the Melnikov function.  As we will see 
in \Sec{sec:algebraic}, the integral of the
flux form over a piece of $\W$ gives flux through that surface to
first order in $\veps$.

  The form $i_{X_{\veps}}\Omega$ has some interesting properties:
\begin{pro}\label{prop:divergence}
   Let $X_{\veps}$ be a perturbation vector field \Eq{eq:pertvect}.  Then
   the form $i_{X_{\veps}}\Omega$ is exact.
\end{pro}

\proof As we already noted,Thm 2.2.24 in \cite{Abraham} implies that 
the divergence of $X_{\veps}$ vanishes, which by definition
$(\div_{\Omega}X_{\veps}) \Omega \equiv L_{X_{\veps}}\Omega$, 
means that the Lie derivative vanishes as well.
Since $d\Omega = 0$ and $L_{X_{\veps}}\Omega
\equiv  d(i_{X_{\veps}}\Omega)+i_{X_{\veps}}d\Omega$, this implies that
$d(i_{X_{\veps}}\Omega) = 0$. Thus the form is closed. Since
$i_{X_{\veps}}\Omega$  is globally defined in  $\R^n$, the form is
exact. \qed

Using this result we can obtain an $(n-2)$-form $\beta$ on $\W$ such that
\begin{equation}\label{eq:represents}
     d\beta \equiv i_{X_{0}}\Omega \;.
\end{equation}
In this case we will say that $\beta$ {\em represents the perturbation} on $\W$.  
Using \Lem{lem:independent} and \Eq{eq:represents}, it is easy to
see that if $M_{\eta}$ exists then
the following $(n-2)$-form is well-defined on $\W$.
\[
      \alpha=\sum_{k=-\infty}^{\infty}(f_{0}^{k})^{*}\beta.
\]
Notice that $\alpha$ is invariant under $f$ and is independent of
$\eta$. Using this we can see that:

\begin{pro}\label{thm:java}
      The flux form $\Phi$ is
      \[
             \Phi =  d\alpha \;.
      \]
\end{pro}
\proof  This is a straightforward calculation using
\Lem{lem:independent} and \Pro{thm:2form}.
\qed

%
%
\subsection{Fundamental Domains}

Our goal in this section is to find a compact subset of the
manifold---a fundamental
domain---that generates the entire manifold under iteration by
$f_{0}$.  We will integrate the flux form over the fundamental domain to show
that the algebraic flux crossing the separatrix is zero. 
From this point on, we will concentrate on the case  $n=3$.
To define the fundamental domain we start with the concept of a proper loop:

\begin{defn}[Proper loop]\label{defn:guasa}
       Let $f_0:\R^{3}\rightarrow\R^{3}$ be a diffeomorphism, and $\W$ a
      forward invariant surface.  We say that a smooth Jordan loop
      $\gamma\subset\W$ is a proper boundary in $\W$ if there $\gamma$ bounds
      a surface $\W_{\gamma} \subset int(\W)$ which is a
      trapping region:
      \[
         f(cl(\W_{\gamma}))\subset int(\W_{\gamma}) \;.
      \]
\end{defn}
Similarly a loop $\gamma$ is a proper loop for a backward invariant
surface if it is a proper loop for the map $f^{-1}$.

It is important to notice that not all invariant surfaces admit
proper boundaries.  A trivial observation is:

\begin{pro}
      If $\gamma$ is a proper boundary in $\W$, then $f(\gamma)$ is
      also a proper boundary.  In addition,
      $\W_{f(\gamma)}=f(\W_{\gamma})$.
\end{pro}

The situation that we have in mind relates to the structure of
stable and unstable manifolds.  Let $a,b$ be compact, normally hyperbolic
invariant sets of $f$, and $\W = W^{s}(a) = W^{u}(b)$ a saddle
connection between them.  A proper loop $\gamma\subset\W$ is a
submanifold of $\W$ that bounds a local submanifold that is a
isolating neighborhood of $a$ in $W^{s}(a)$.  In other words
$\gamma$ is proper if it bounds an open local submanifold,
$W_{loc}^{s}(a) = \W_{\gamma}$, that maps inside itself.

If $\gamma$ is proper, we can define the stable manifold {\em starting
at} $\gamma$, denoted by $\W_{\gamma}=W_{\gamma} ^{s}(a)$, as the
closure in $W^{s}(a)$ of the local stable manifold bounded by
$\gamma$.  In the same way, for $b$, if we have a proper loop $\sigma$
for $f^{-1}$, we define the unstable manifold {\em up to} $\sigma$,
denoted $W_{\sigma}^{u}(b)$, as the interior of the local unstable
manifold bounded by $\sigma$.  We will see below why it is convenient
to use this slightly asymmetric definition.

Given a proper loop  we can define

\begin{defn}[Fundamental domain]
      Let $\W$ be a forward invariant surface.  An
      submanifold with boundary, $\cP$, is a fundamental domain of
      $\W$ if there exists some proper loop $\gamma$ in $\W$, such that
      \[
         \cP=\cP_{\gamma}=\W_{\gamma}\setminus\W_{f(\gamma)} \;.
      \]
\end{defn}
The fundamental domain is a manifold with the boundary
\[
     \partial\cP=\gamma\cup f(\gamma) \;,
\]
see \Fig{fig:fundamental}. An immediate consequence of the definition is 
that all the forward iterations of a fundamental domain are also 
fundamental and $\cP_{f(\gamma)}=f(\cP_{\gamma})$.  It is easy to see that, if
proper boundaries exist, then the forward invariant manifold can be decomposed
as the disjoint union of fundamental domains.
\[
      \W=\left(  \W\setminus\W_{\gamma}\right)
      \cup\bigcup_{k\geq0}f^{k}\left(  \cP\right)  \;.
\]
If the surface $\W$ is both forward and backward invariant, then this
decomposition works in both directions. In such case we have
\[
      \W=\bigcup_{k\in{\mathbb{Z}}}f^{k}\left(  \cP\right)  \;.
\]

\InsertFig{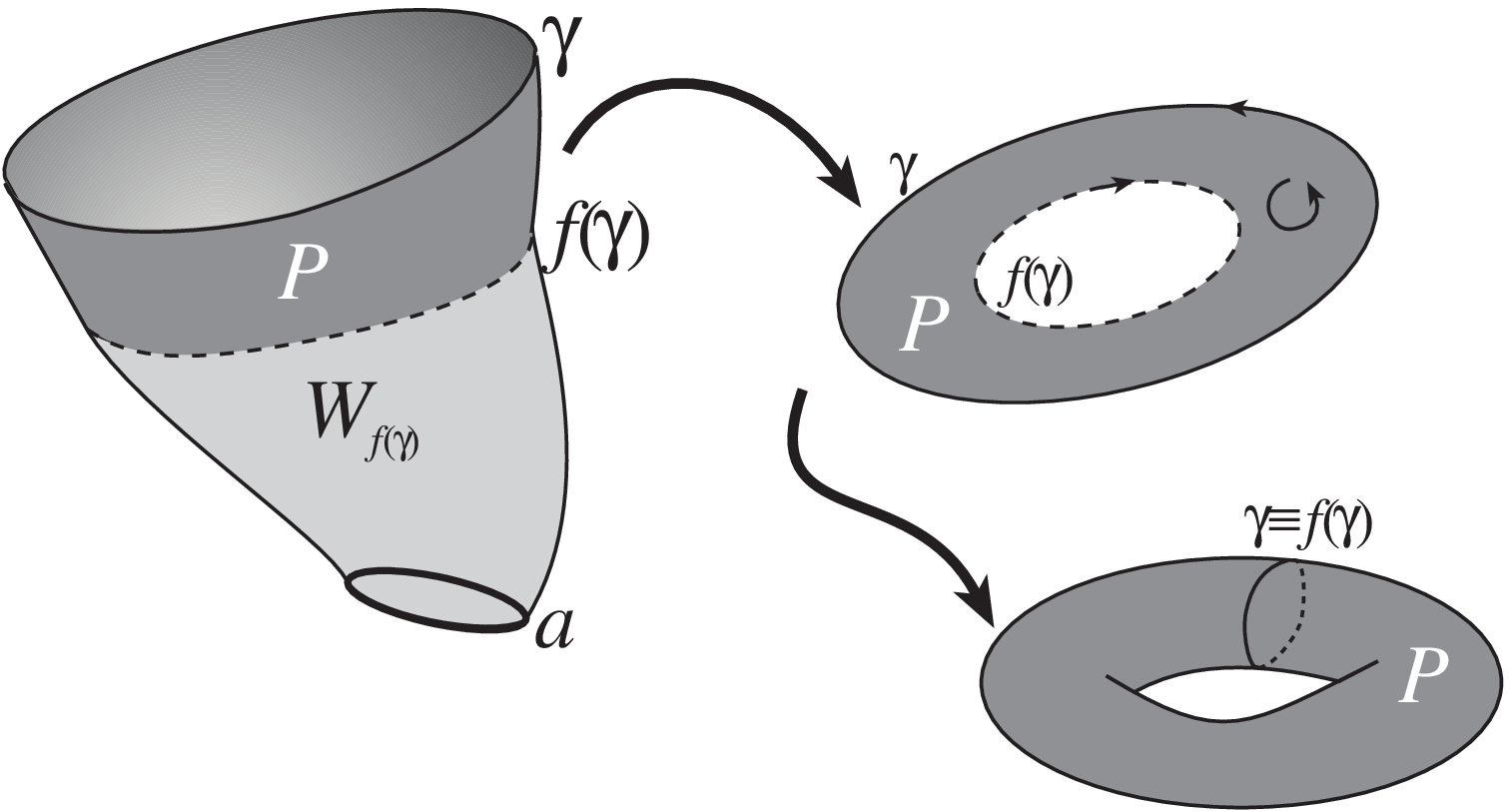}
{Fundamental domain $\cP$ on a stable manifold of an invariant circle $a$
bounded by a loop $\gamma$ and its image $f(\gamma)$.
The second part of the figure shows the annular fundamental domain itself,
together with an assigned orientation. Finally, if we identify the points on
$\gamma$ with their images, then the fundamental domain is equivalent to a torus.}
{fig:fundamental}{5in}

%
%
\subsection{Algebraic Flux}\label{sec:algebraic}

Given a vector field $X$, the differential form $i_{X}\Omega$ represents the flux associated with $X$; that is given set of vectors $v_{1},v_{2},...v_{n-1}$, $\Omega(X,v_{1},v_{2},...v_{n-1})$  is the volume of the parallelepiped formed from these vectors, and thus measures the rate at which volume is swept out by $X$ through the  parallelepiped defined by $v_{1},v_{2},...v_{n-1}$.

According to \Eq{eq:fluxform}, the form $\Phi$ is the sum of $i_{X_{0}}\Omega$ along an orbit on $\W$. Thus 
$\Phi$ evaluated at a point on a fundamental domain $\cP$ measures the total flux of 
$X_{0}$ along the orbit of that point.

The algebraic flux through a surface is the integral of the flux over the surface. Since $\Phi$ measures the flux along an orbit on $\W$, the integral of $\Phi$ over a fundamental domain is the algebraic flux through the entire surface $\W$.

\begin{pro}\label{thm:ZeroAlgebraicFlux}
      The algebraic flux through $\W$ is zero: 
          $\int_\W i_{X_{0}}\Omega = \int_{\cP} \Phi=0$.
\end{pro}

\proof The fundamental domain $\cP=\cP_{\gamma}$ is a submanifold with
boundary, such that $\partial\cP_{\gamma}=\gamma\cup f(\gamma)$,
where $\gamma$ is closed curve that does not intersect $f(\gamma)$.
If we give an orientation $\left[ \cP\right] $ to $\cP$, the induced
orientation on the boundary satisfies $\left[ \gamma\right] =-\left[
f(\gamma)\right] $, recall \Fig{fig:fundamental}.
Since $\Phi = d\alpha$ by \Pro{thm:java}, and $\alpha$ is invariant under $f$, 
Stokes's theorem, implies
\begin{align*}
       \int_{\cP} \Phi
             &  =\int_{\cP}d\alpha =\int_{\partial\cP}\alpha
                  =\int_{\gamma}\alpha+\int_{f(\gamma)}\alpha\\
             &  =\int_{\gamma}\alpha-\int_{\gamma}f^{*}\alpha=0 \; \qed
\end{align*}

A simple corollary is 
\begin{cor}
      The Melnikov function must have zeros on $\cP$.
\end{cor}

The importance of fundamental domains is that much of the information
about the entire manifold can be found by looking only at these
submanifolds.  In particular, if we have a transversal intersection of
two invariant surfaces, we can look at a pair of fundamental domains
and study primary intersections.  In \Fig{fig:dostres} we show a pair
of fundamental domains of two stable and unstable manifolds that
intersect transversally.

\InsertFig{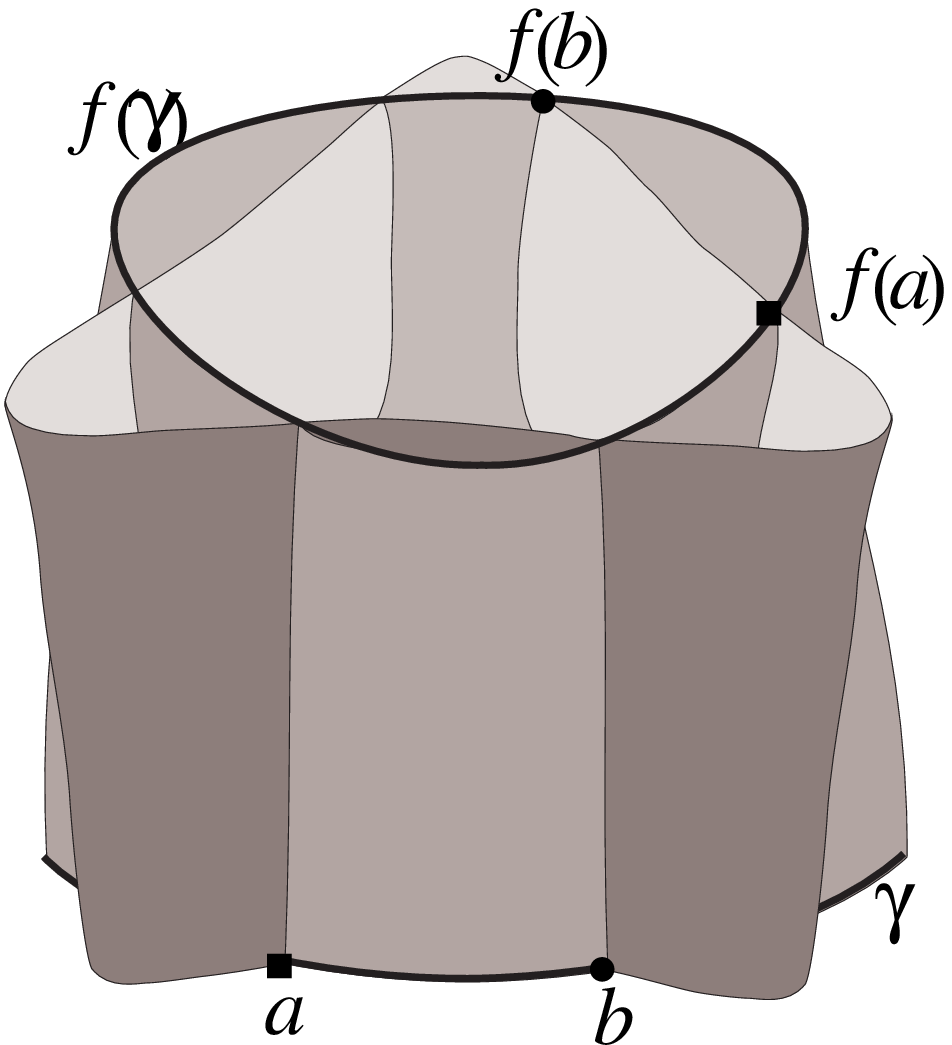} {A pair of fundamental domains for two
       different circles $C_{1}$ and $C_{2}$ that intersect
       transversally, forming three-dimensional lobes.} {fig:dostres}{3in}

The curves of zeros of the Melnikov function can be classified by their homology on $\cP$. To do this, we identify two boundaries of the fundamental annulus by identifying $\gamma$ with $f(\gamma)$. With this identification the fundamental annulus becomes a torus, as sketched in \Fig{fig:fundamental}. Since the homology group of the torus is $\Z^{2}$, we can label the curves by a pair of integers $(m,n)$ which represent the number of times 
the curves wrap around each circuit of the torus. For example when the identification is performed on  \Fig{fig:dostres} there are a pair of zero crossing curves with homology type $(3,1)$---they move once around the annulus in three vertical transits.  

\section{Examples} \label{sec:examples}

In this section we construct a family of volume-preserving maps that
have a saddle connection between a pair of invariant circles.  We
obtain this family by starting with an area-preserving twist map that
preserves an axis and extending it to a three-dimensional,
volume-preserving map by composing it with a sheared rotation about
that axis.  The twist map is defined in such a way that it has a
saddle connection between two fixed points, and so the resulting
three-dimensional map has a pair of invariant circles with a
two-dimensional connection.  Examples similar to these were found by
Lomel\'{\i} \cite{Lomeli96} and are closely related to those in
\cite{Lomeli00}.

We begin with
an area-preserving map on $\R^2$ in coordinates $(z,r)$ that preserves
the axis $r=0$, and has a fixed point at some nonzero $r=r^{*}$.  For
example, set
\[
      (r',z') = G(r,z) = \left(h^{-1}(r+h(z))-z ,\; h(z) + r- r^{*} \right) \;,
\]
where $r^{*} \in \Z$.  Here we assume that $h: \R \rightarrow \R$
is an increasing circle diffeomorphism of period $1$, i.e.,
$h(z+1)=h(z)+1$.  Moreover we can verify that $\det(DG) = 1$, so
that $G$ is area-preserving.  Finally
\[
      G^{-1}(r,z) = \left( z-h(h^{-1}(z)-r) ,\; h^{-1}(z)+r^{*}-r \right) \;,
\]
so that $G$ is a diffeomorphism.

It is easy to see that $G(0,z) = (0,\;h(z)-r^{*})$ so that the $z$-axis is
preserved. The map has fixed points at solutions of $z = \frac12 (
h^{-1}(z) + h(z))$, with $r = r^{*} + z - h(z)$.  In particular, any
hyperbolic fixed point of $h$, $z^{*} = h(z^{*})$, yields a saddle fixed point
$(r^{*},z^{*})$ of $G$ whose multipliers are $\lambda = h'(z^{*})$ and
$1/\lambda$.  Between every pair of such fixed points of $h$ there is
at least one other fixed point of $G$; it is typically elliptic.

The map $G$ is not necessarily integrable (in \Sec{sec:integrable}
we will choose an $h$ that leads to an integrable map). However, $G$ always
has a pair of invariant curves:
\begin{align} \label{eq:invariants}
     \W_{0} &=\left\{ (z,r):  r = r^{*}\right\}  \;, \nonumber \\
     \W_{1} &=\left\{(z,r):  r = h^{-1}(z)-h(z)+r^{*}\right\} \;.
\end{align}
These curves intersect at any fixed point $z^{*}$ of $h$.  Thus they
provide a saddle connection between points
$(z_{1}^{*}, r^{*})$ and $(z_{2}^{*}, r^{*})$, where $z^*_i$ are neighboring
fixed points of $h$.

We show an example of the dynamics of $G$ in \Fig{fig:arnold} for the
case that $h(z) = z-\frac{k}{2\pi} \cos (2\pi z)$, and $r^{*} = 1.0$.
Here one can see the saddle connection at $r=r^{*}$ as well as chaotic
dynamics in other regions of phase space.

\InsertFig{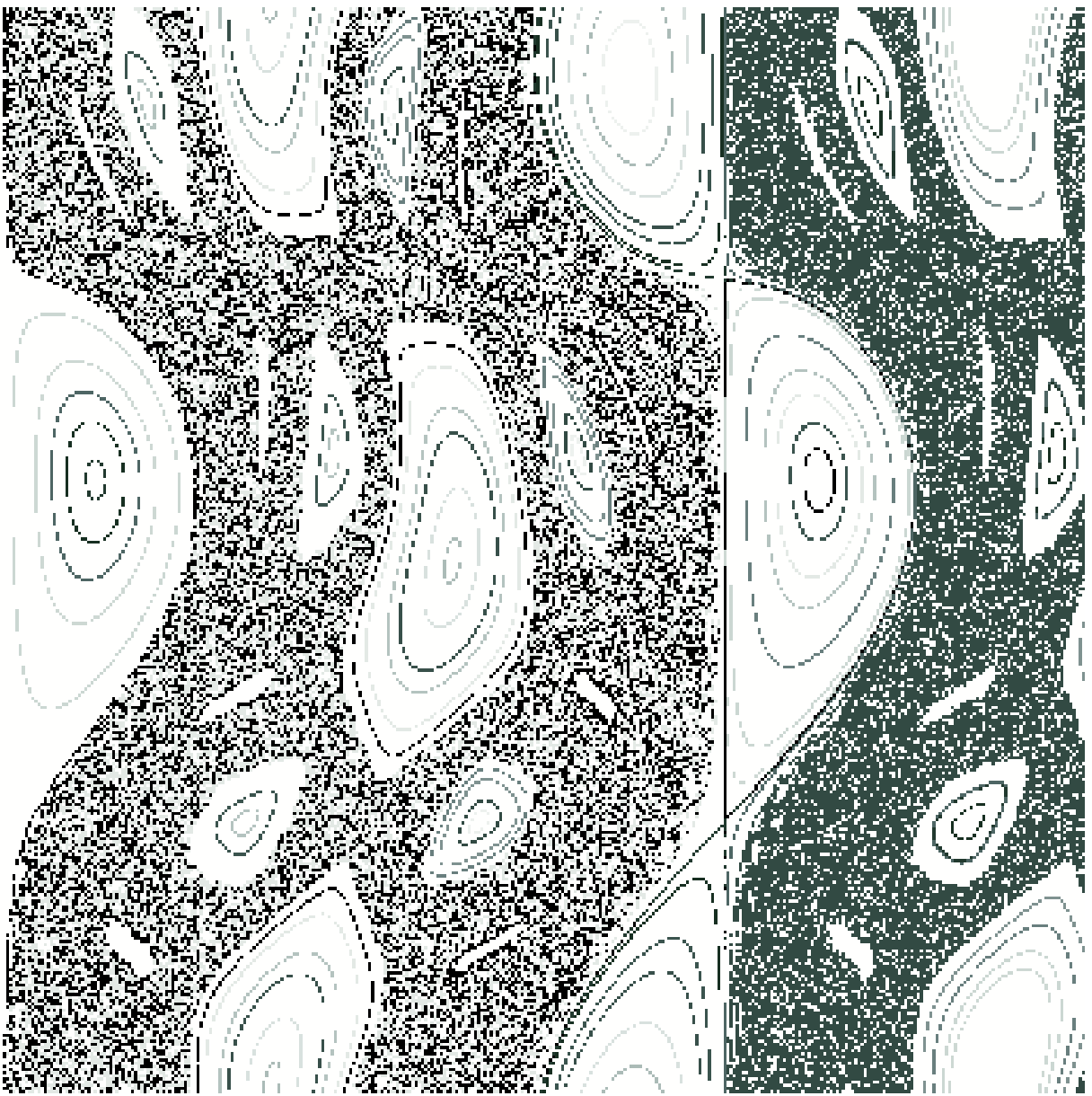} { Dynamics of $G$ for the Arnold circle map
with $k=0.9$. The domain of the figure is $[0,1.5]\times [-0.5,0.5] $.}
{fig:arnold}{4in}

We can extend $G$ to $\R^3$ by introducing the cylindrical angle
$\theta$ and using the volume-form $\Omega = dr \wedge d\theta
\wedge dz$.  Defining $r$ to be the ``symplectic" polar radius,
\begin{equation}\label{eq:radius}
     r=\frac{1}{2}(x^{2}+y^{2}) \;,
\end{equation}
the cylindrical coordinates are
\[
       (x,y,z) = P(r,\theta,z)= (\sqrt{2r} \cos \theta , \sqrt{2r} \sin
\theta, z)
\]
so that $\Omega = dx \wedge dy \wedge dz$.  In terms of these
coordinates the map becomes $g = P\circ G \circ P^{-1}$:
\[
       g(x,y,z)=\left(\rho(r,z)x ,\; \rho(r,z)y, \;r+h(z)-r^{*} \right)\;,
\]
where $\rho = \sqrt{r'/r}$ is explicitly
\begin{equation} \label{eq:rho}
       \rho(r,z)=\left\{
       \begin{array} [c]{cc}
      \displaystyle\sqrt{\frac{h^{-1}(r+h(z))-z}{r}}, & r\neq0\\
      & \\
      \left(h^{\prime}(z)\right)^{-\frac12}, & r=0.
       \end{array}
       \right.
\end{equation}

It was shown in \cite{Lomeli00} that if $h$ is $C^{r}$, then
$\rho$ is $C^{r-1}$ so that, in this case, $g$ is a diffeomorphism.

The map becomes fully three-dimensional if we introduce dynamics in
the angle $\theta$.  To do this, we compose the map with a rotation.
Denote  the rotation about the $z$-axis by angle $\psi$ by
\begin{equation} \label{eq:rotmatrix}
      R_{\psi}=\left(
      \begin{array}[c]{ccc}
     \cos\psi & -\sin\psi & 0\\
     \sin\psi & \cos\psi & 0\\
     0 & 0 & 1
      \end{array}
      \right)  \;.
\end{equation}
Introducing a rotation angle $\tau(r,z)$ that depends smoothly on
$(r,z)$, we define a diffeomorphism $f$ by
\begin{equation}\label{eq:intmap}
      f=g\circ R_{\tau} \;,
\end{equation}
Note that since $R_{\tau}$ and $g$ both preserve $\Omega$, so does
$f$.

The map $f$ still has a rotational symmetry
\begin{equation}\label{eq:rotsymm}
      f\circ R_{\psi}=R_{\psi}\circ f \; ,
\end{equation}
for any constant $\psi\in\R$. This implies that when $G$ has a
saddle connection, so does $f$:

\begin{pro}
      The surfaces \Eq{eq:invariants} are invariant under \Eq{eq:intmap}.
      In addition, $\W_{0}$ and $\W_{1}$ intersect on the invariant circles
      \[
     \C (z^{*}) =\left\{ (x,y,z):z=z^{*}, r = r^{*}\right\} \;,
      \]
      where $z^{*}$ is any fixed point of $h$.
\end{pro}

Every point on the circles $\C (z^{*})$ is fixed
under $g$. The derivative of $g$ at such points is
\begin{equation}\label{eq:derivg}
     Dg(x,y,z)=
     \left(
       \begin{array} [c]{ccc}
         \frac{1}{2r^{*}}(\lambda^{-1}x^2 +y^2)
        &\frac{1}{2r^{*}}(\lambda^{-1}-1)xy    & 0\\
        &  & \\
         \frac{1}{2r^{*}}(\lambda^{-1}-1)xy
        &\frac{1}{2r^{*}}(x^2+\lambda^{-1}y^2) & 0\\
        &  & \\
         x & y & \lambda
      \end{array}
      \right) \;,
\end{equation}
where $\lambda=h^{\prime}(z^{*})$.  More generally, we can compute the
derivative of $f^{n}$ on the invariant circles for the special case
that the rotation angle is constant:

\begin{pro}
       Suppose $f$ is given by  \Eq{eq:intmap}, that $\tau$ is 
       constant, $z^{*} = h(z^{*})$,and $\C (z^{*})$ is the 
       corresponding invariant circle.  Then for all
       $(x,y,z)\in\C (z^{*})$
       \begin{equation} \label{eq:deriv}
        Df^n (x,y,z)=R_{n\tau}\left(
         \begin{array} [c]{ccc}
            \frac{1}{2r^{*}}(\lambda^{-n}x^2+y^2) 
           &\frac{1}{2r^{*}}(\lambda^{-n} -1) xy   & 0\\
           &  & \\
            \frac{1}{2r^{*}}(\lambda^{-n}-1)  xy   
           &\frac{1}{2r^{*}}(x^2+ \lambda^{-n} y^2) & 0\\
           &  & \\
            \displaystyle\frac{(\lambda^{2n}-1)\,x}{\lambda^{n-1}(\lambda^2-1)} 
           &\displaystyle\frac{(\lambda^{2n}-1)\,y}{\lambda^{n-1}(\lambda^2-1)} 
           &\lambda^n
        \end{array}\right)  \;,
       \end{equation}
      where $\lambda=h^{\prime}(z^{*})$.  Moreover, if
      $\lambda > 1$ $(< 1)$ the invariant circle has a stable
      (unstable) manifold contained in $\W_{1}$, and unstable (stable)
      manifold contained in $\W _{0}$.
\end{pro}

\proof Given the symmetry \Eq{eq:rotsymm}, it is enough to check
\Eq{eq:deriv} for points of the form $$(\sqrt{2r^{*}},0,z^{*}).$$
Since \Eq{eq:deriv} reduces to \Eq{eq:derivg} when $n=1$, it is
enough to verify the induction step
\[
      Dg(\sqrt{2r^{*}},0,z^{*})Dg^n(\sqrt{2r^{*}},0,z^{*})=
         Dg^{n+1}(\sqrt{2r^{*}},0,z^{*}) \;.
\]
The vector $(0,0,1)^{t}$ is an eigenvector of $Dg(x,y,z^{*})$ with
eigenvalue $\lambda$.  Since this is tangent to $\W_0$, this shows
that it is the stable manifold when $\lambda < 1$.  Similarly the
vector $(x, y ,2r^{*}\frac{\lambda}{1-\lambda^2})^{t}$ is an
eigenvector with eigenvalue $\lambda^{-1}$ that is tangent to $\W_1$.
The final eigenvector of $Dg$ is $(y,-x,0)^{t}$ which is tangent to
$\C$ and has eigenvalue $1$.  \qed

\subsection{Integrable case}\label{sec:integrable}
In general the maps $g$ and $f$ are not integrable, even though they
have a saddle connection.  However, for a special choice of $h$ there
is an integral.  This example is related to the work of Suris
\cite{Suris1, Suris2} on area-preserving integrable maps, but is
distinct from the three-dimensional maps found in \cite{Gomez02} that
have an invariant but which do not have a rotational symmetry.

Let $m(w)= \frac{aw+b}{cw+d}$ be the M\"{o}bius transformation on
$\R\cup \{\infty\}$ with $ad-bc = 1$.  A circle map conjugate to
$m$ is obtained by defining $w= \tan \pi z$, giving $h$
\begin{equation}
       h_{m}(z)=\frac{1}{\pi}\arctan\left( m(\tan \pi z) \right) \;.
\end{equation}
This map can be written more explicitly as a circle map using
trigonometric identities:
\begin{equation}\label{eq:hM}
      h_{m}(z) = z +\frac{1}{\pi}\arctan\left[
                    \frac{b-c + (b+c)\cos(2\pi z)+(a-d)\sin(2\pi z)}
                    {a+d - (a-d)\cos(2\pi z)+(b+c)\sin(2\pi z)}
                  \right]
\end{equation}

\InsertFig{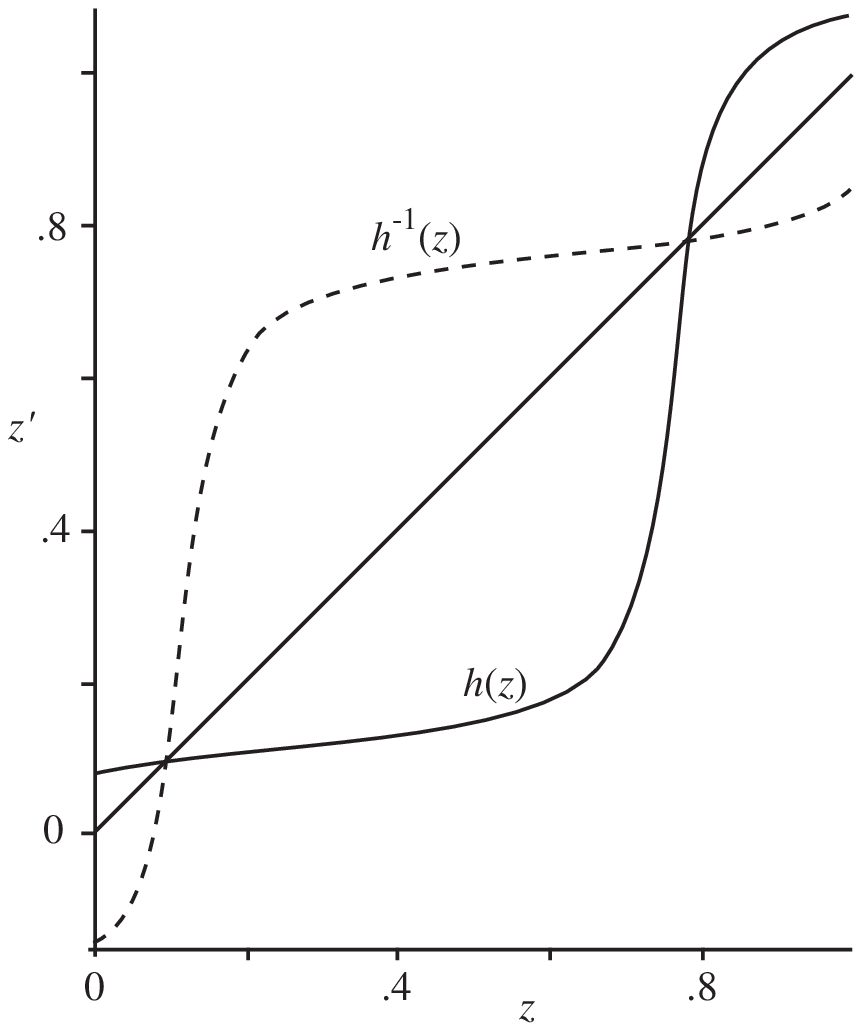} {The circle map $h_{m}(z)$ and its inverse
for $(a,b,c,d) = (1,\frac12,2,2)$.}{fig:circlemap}{3in}

Requiring $ad-bc = 1$, some useful properties
of this family of circle maps follow easily from its conjugacy to the M\"{o}bius
transformation
\begin{itemize}
     \item $h_{m}^t = h_{m^t}$, for all $t\in \Z$;
     \item If $|\tr(m)| >2$ then $h_{m}$ has two fixed points $z_{\pm}^{*}
     \in [0,1)$.
     \item The fixed points have multipliers $Dh(z_{\pm}^{*}) =
     \frac{4}{(\sigma\pm \sqrt{\sigma^{2}-4})^{2}} > 0$, where $\sigma=a+d$.
     \item Thus $z_{-}^{*}$ is unstable and $z_{+}^{*}$ is stable.
\end{itemize}

For any $h_{m}$, the resulting map $G$ is integrable. To see this, we show how
this map is related to the Suris example. First note that $G$ can be
rewritten as a second difference equation
\[
         z_{t+1} -2z_{t}+z_{t-1}= h(z_{t})+h^{-1}(z_{t})-2z_{t}=F(z_{t}) \;,
\]
where $r_{t}= z_{t+1}-h(z_{t})+r^{*}$.  Suris showed that this family
is integrable when $F$ is given by
\[
     F(z) = \frac{1}{\pi}\arctan\left[ \frac{A\sin(2\pi x) +B\cos(2\pi x) +
                 C\sin(4\pi x)+D\cos(4\pi x)}{1-E-A\cos(2\pi x) +B\sin(2\pi x)
            -C\cos(4\pi x)+D\sin(4\pi x)}\right] \;,
\]
for any values of the parameters $A,B,C,D,E$.  After some algebra one can see
that our map has this form with $A = b^{2}-c^{2}$, $B=(a-d)(c-b)$,
$C=\frac12(b+c)^{2}-\frac12(a-d)^{2}$, $D = (d-a)(b+c)$, $E =
\frac12(a^{2}+b^{2}+c^{2}+d^{2})$.

The map $G$ with this $h$ has the integral
\begin{align*}
     J(z,z') = &(1-E)\cos(2\pi(z-z')) - A(\cos(2\pi z)+\cos(2\pi z')) +\\
                  &\quad B(\sin(2\pi z)+\sin(2\pi z'))
                    - C\cos(2\pi (z+z'))+D\sin(2\pi (z+z')) \;.
\end{align*}

For the examples, we will use the map \Eq{eq:intmap} with 
$h_{m}$ given by \Eq{eq:hM} with
\begin{equation}\label{eq:Mnu}
     m(w) = \frac{(\nu+1)w+ \nu -1}{(\nu -1)w +\nu+1}
\end{equation}
This corresponds to setting $m$ to the hyperbolic
rotation matrix $a=d= \cosh(\ln(\nu))$, $b=c= \sinh(\ln(\nu)) $.
In this case $m^{t}$ is given by replacing $\nu$
with $\nu^{t}$; thus iteration of $h$ is extremely easy.  This was
also the example used in \cite{Lomeli00}.

This gives a family of three-dimensional maps, $f$, with parameters $\nu$\
and $\tau$. Setting $r^* = 1$, there are invariant circles at 
$(r^*,z^*) = (1,\pm \frac14)$.
For this case the invariant has the form
\begin{equation}\label{eq:J}
      J(x,y,z) = 2\nu \cos(2\pi r) + (1-\nu^2)\cos(2\pi z)\sin(2\pi r) \;.
\end{equation}
The level sets corresponding to $J= 2\nu $ give the invariant
manifolds $\W_0$, $\W_1$, and the circles $\C(z^{*})$.  The level sets
of $J$ are shown in \Fig{fig:levelsets} for the case $\nu=0.3$.
Though the level sets of $J$ make it appear
that  $(\frac12,\pm\frac14)$  are also 
invariant; $G(\frac12,\frac14) = (\frac12,-\frac14)$, so these points 
move downward. Moreover the curve $\{ r = \frac12\}$ has 
image $\{ r = h^{-1}_{\nu}(z)-h_{\nu}(z) + \frac12 \}$, 
and this latter curve has image again of $r = \frac12$.

\InsertFig{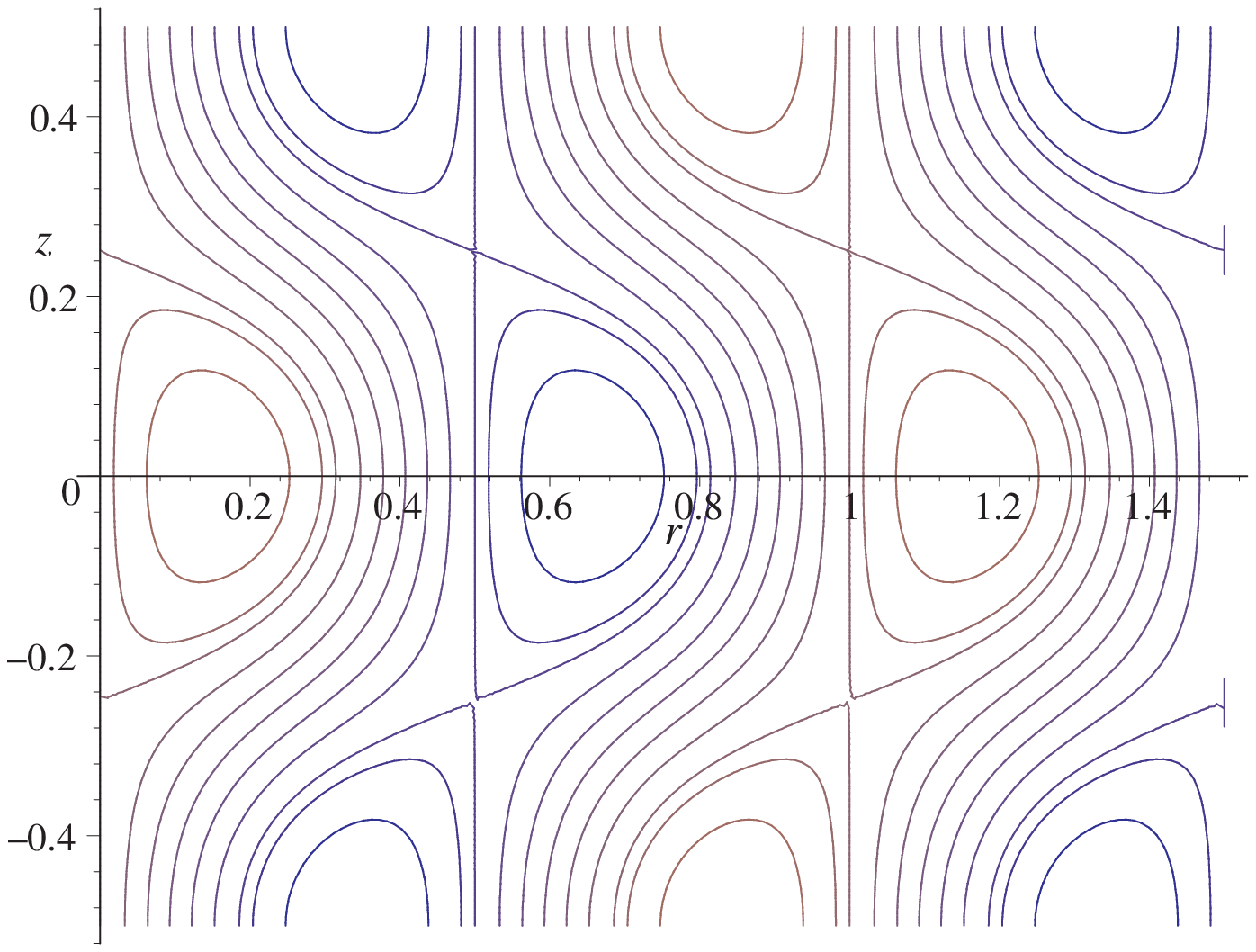} {Levels sets of $J$ for $\nu=0.3$ Here the
maximum of $J$ occurs at $z=0$ near $r=1.1$, while a minimum
occurs near $r=0.7$.}{fig:levelsets}{4in}

\subsection{Perturbed Map}
We break the invariant surfaces by choosing a perturbation of the
form \Eq{eq:pert}.  For the first example, we choose a composition
of two simple perturbations:
\begin{equation}\label{eq:pert1}
  \begin{split}
     P_1(x,y,z) &=  \left((1+y^2)({z^{*}}^2-z^2),0,0\right) \;, \\
     P_2(x,y,z) &= \left(0, x({z^{*}}^2-z^2),0\right) \;. \\
  \end{split}
\end{equation}
Each of these maps has a nilpotent Jacobian, which implies that the
maps $id +\veps P_i$ are volume-preserving for all $\veps$.  The
complete perturbation is then defined as
\[
      id + \veps  P_\veps =  (id + \veps  P_2)\circ (id + \veps  P_1) \;.
\]
Substituting the perturbation into the computation \Eq{eq:pertvect} for the vector
field $X_\veps $ gives
\[
      X_0 =  P_2+ P_1 \;.
\]

Fundamental domains $\cP_{i}$ on $\W_{i}$ are given by the
annuli bounded by the circles $\gamma_{i} = \{z=0\}\cap \W_{i}$ and
their images, $f(\gamma_{0}) = \{z=h(0)\}\cap \W_{0}$ or
$f(\gamma_{1})=\{z=h^{-1}(0)\}\cap \W_{1}$, respectively.  These
 can be projected onto $(z,\theta)$
coordinates for visualization.  Calculation of the Melnikov sum
\Eq{eq:melnikovsum} is straightforward using the adapted form $dJ$
associated with the invariant \Eq{eq:J}.

\InsertFig{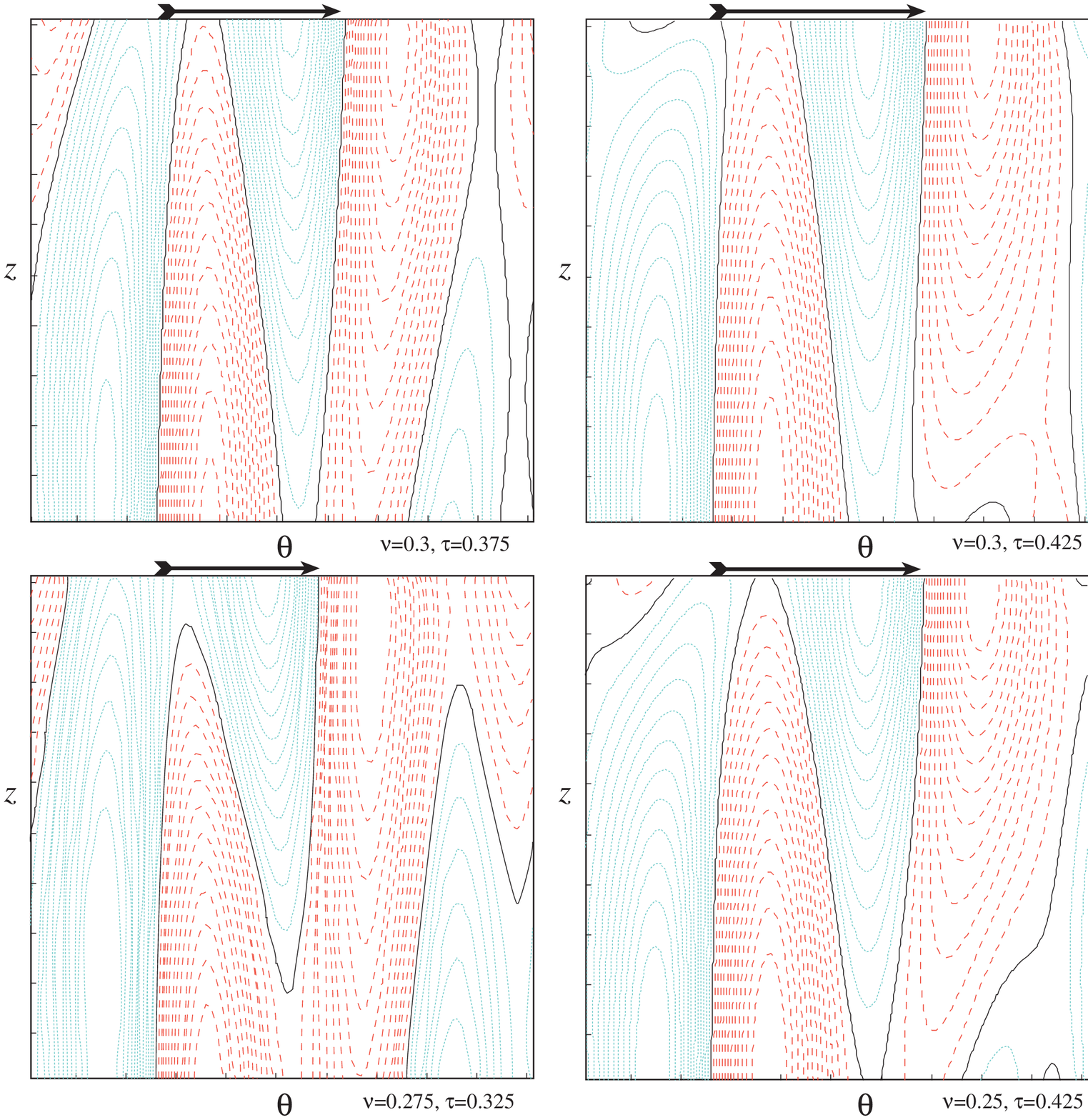} {Contours of the Melnikov function for $\W_{0}$
with the perturbation $\Eq{eq:pert1}$.  Shown are four values of the
parameters $(\nu,\tau)$.  Bounds for the figures are
$z=[0,h^{-1}(0)]$ and $\theta=[0,2\pi]$. The arrow at the top of each 
panel shows the translation by $\tau$.} {fig:contours}{5in}

We show several representative contour plots of $M_{dJ}$ in
\Fig{fig:contours}.  In the figure, positive values of $M_{dJ}$ are shown
as dashed lines and negative as dotted lines, while the zero contour
is the solid line.  For example, in the bottom-left panel ($\nu=0.275$
and $\tau=0.325$) there are two zero contours, corresponding to the
unstable and stable manifolds crossing with opposite signatures (since
the algebraic flux through the fundamental annulus is zero, the zero
contours must come in pairs).  Since the unperturbed map takes the
circle $\gamma_{1}$ to the circle $f(\gamma_{1})$ shifting each point
by $\tau$, the lower boundary of $\cP$ can be identified with the
upper boundary after shifting the latter to the right by $\tau$
(we show this shift by the arrows in \Fig{fig:contours}).  After this
identification the fundamental annulus becomes a torus, and the zero
contours correspond to a pair of circles that wrap once vertically.
Thus these contours have homology type $(1,0)$.  The bottom right
panel also has this homology type, though the curves are very close to
a bifurcation.

There are several such bifurcations in homology type of the zero
contours as we vary the parameters.  For example in the upper left
panel, the homology type is $(3,1)$--- as each zero contour moves from the bottom to the top of $\cP$, it lags the maps translation of $\theta$ by a full circuit in three vertical transits.  In the top-right panel there are two zero contours with
the homology $(2,1)$.  To elucidate these changes in homology, we show
a bifurcation diagram in the space of the parameters in
\Fig{fig:fluxbif1}.  We have only found the three homology classes
already mentioned.

Also shown in \Fig{fig:fluxbif1} are contours of the geometric flux
\[
     \mbox{Flux} = \frac12 \int_{\cP} | \Phi |
\]
as a function of
$\nu$ and $\tau$. The flux is largest when $\nu$ and $\tau$ are both small,
and it appears to get extremely small as $\nu$ approaches one. Note that
there is a ``valley'' in the flux contours near both homology bifurcations.

\InsertFig{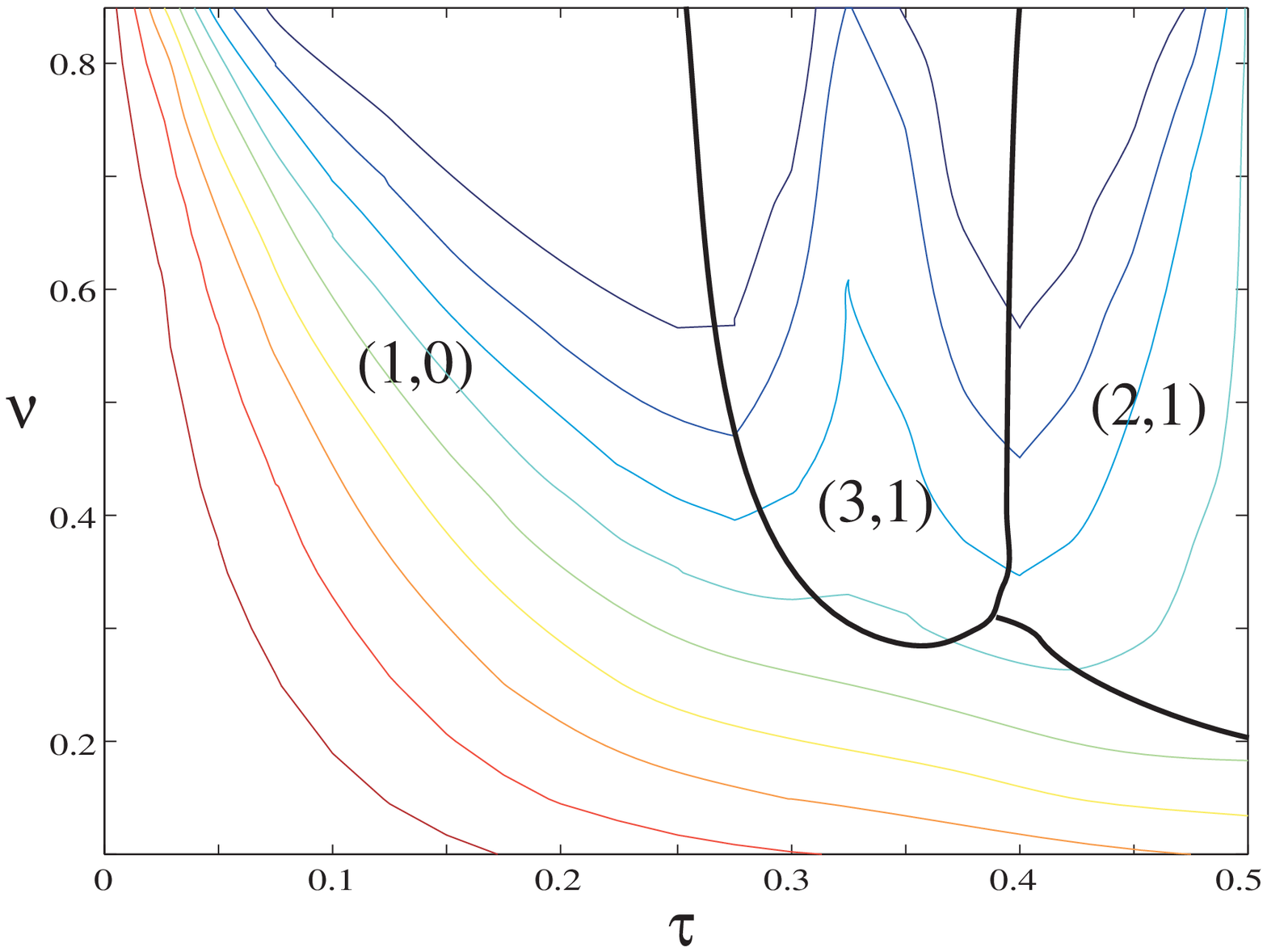} {Contours of the geometric flux through $\W_{0}$
as a function of $\nu$ and $\tau$ for the perturbation
\Eq{eq:pert1}.  The nine contours are at equally spaced levels ranging
from a flux of $0.09$ at the lower left to $0.01$ at the top.  Also shown are
bifurcation curves corresponding to the change in homology types of
the zero contours of $M_{dJ}$.} {fig:fluxbif1}{3in}

Finally, we have also studied the perturbation
\begin{equation}\label{eq:pert2}
   \begin{split}
     P_1(x,y,z) &=  \left((1+y^2)({z^{*}}^2-z^2),0,0\right) \;, \\
     P_2(x,y,z) &= \left(0, x^2({z^{*}}^2-z^2),0\right) \;, \\
     P_3(x,y,z) &=  \left(0,0,r-r^{*}\right) \;.
   \end{split}
\end{equation}
giving a perturbation vector field $X_{0}=P_{1}+P_{2}+P_{3}$. We
show the bifurcation diagram for the zero contours of $M_{dJ}$ for
$\W_{0}$ in \Fig{fig:fluxbif2}. For this case there appear to be
only two homology types, $(1,0)$ and $(3,1)$. Again there is a
``valley'' in the flux near the bifurcation curve.

\InsertFig{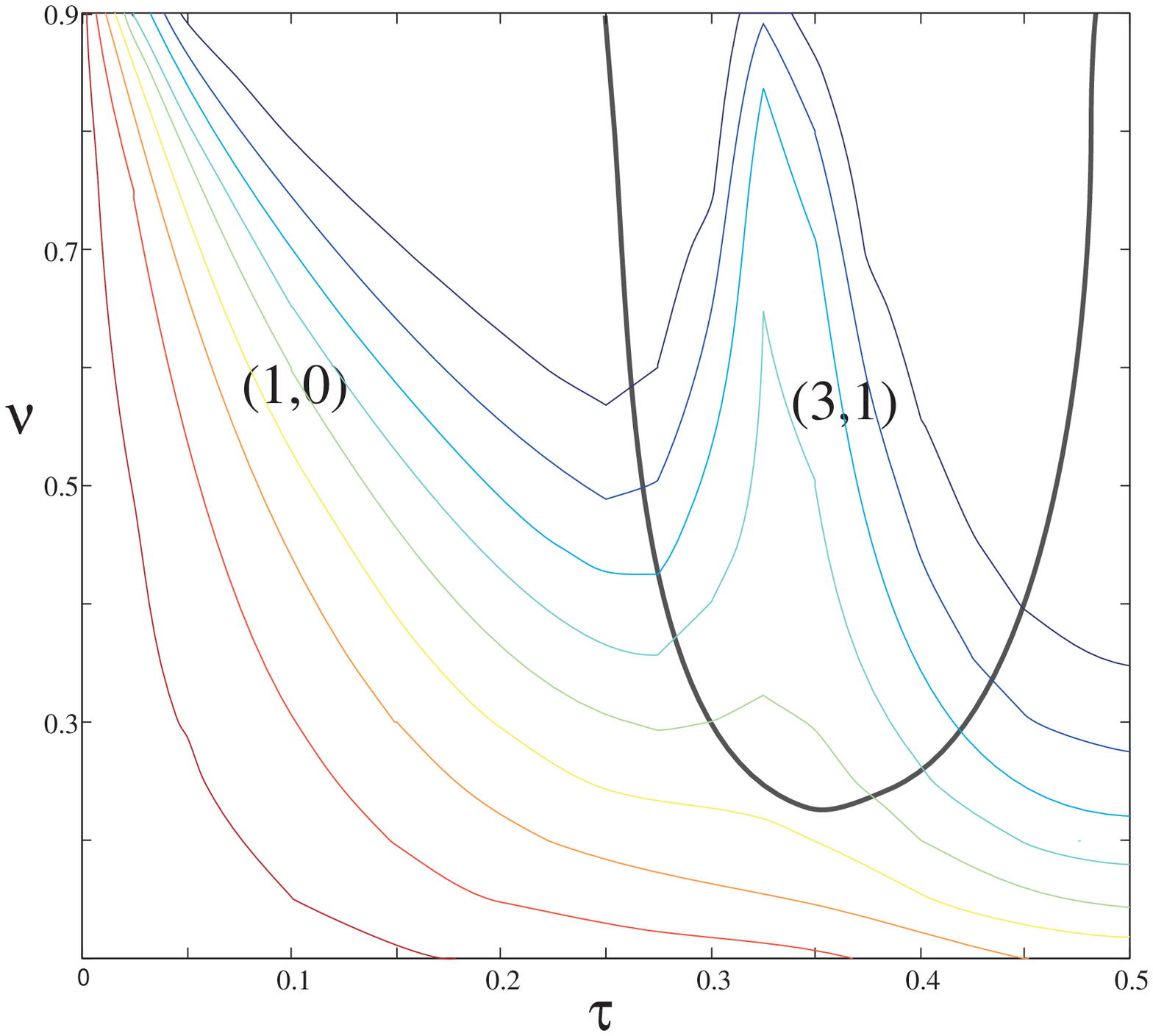} {Contours of the geometric flux through $\W_{0}$
as a function of $\nu$ and $\tau$ for the perturbation
\Eq{eq:pert2}.  Also shown are bifurcation curves corresponding to the
change in homology types of the zero contours of $M_{dJ}$.  In
this case there is only one curve of bifurcation, corresponding to
$(1,0) \longleftrightarrow (3,1)$.} {fig:fluxbif2}{3in}

We have also computed the Melnikov function for the second invariant
set, $\W_{1}$, but do not show the curves since they are very similar
to those for $\W_{0}$.

\section{Conclusion}
We have shown that the flux-form $\Phi$ is the unique $(n-1)$-form on a
codimension-one saddle connection that describes the lowest order splitting 
of the manifolds upon perturbation. The integral of the one-way flux 
over a fundamental domain characterizes the transport across the manifolds 
in the perturbed system. For our example, the magnitude of the flux is strongly
correlated with bifurcations in the homology of the crossing curves---near 
a bifurcation the flux is small. It would be nice to understand if this is 
a general feature of transport for volume-preserving maps.

In the future we also hope to study the evolution of the full manifolds 
numerically, to compare with our Melnikov results. We would also 
like to develop a nonperturbative method to compute the geometric 
flux, analogous to the action techniques for symplectic maps 
\cite{Meiss92}. With this, we would like to verify that the geometric flux 
quantifies the transport observed numerically.

\bibliographystyle{unsrt}
\bibliography{circle}
\clearpage
\end{document}